\pdfoutput=1
\documentclass{IEEEtran}
\usepackage[utf8]{inputenc}
\usepackage[ruled,vlined]{algorithm2e}
\usepackage{xspace}
\usepackage{graphicx,
	psfrag,
	epsfig,
	epstopdf,
	amsthm,
	amssymb,
	url,
	subcaption,
	balance,
	enumerate,
	color,
    url,
	setspace,
	tikz,
	pgfplots
}
\usepackage{amsmath}
\usepackage[nospace,noadjust]{cite}
\usepackage{pbox}
\usepackage{geometry}
\usepackage{multirow}
\usepackage{adjustbox}
\usepackage{array}
\usepackage{booktabs}
\usepackage{float}
\usepackage{stfloats}
\usepackage{multicol}
\usepackage{pbox}
\usepackage{booktabs}
\usepackage{tabularx}
\usepackage{xcolor}
\usepackage[normalem]{ulem}

\newcommand{\cref}[1]{Constraint~\ref{#1}}

\newcommand{\ignore}[1]{}
\newgeometry{left=1.2cm,right=1.2cm,bottom=1.5cm,top=1.5cm}

\begin{document}

\title{Five-Layers SDP-Based Hierarchical Security Paradigm for Multi-access Edge Computing}

	\author{
	\IEEEauthorblockN{Jaspreet Singh \IEEEauthorrefmark{1}, Yahuza Bello \IEEEauthorrefmark{1}, Ahmed Refaey \IEEEauthorrefmark{1}\IEEEauthorrefmark{2}, and Amr Mohamed \IEEEauthorrefmark{3} \thanks{Corresponding author: Ahmed Refaey} \thanks{email: ahmed.hussein@manhattan.edu}}\\

	\IEEEauthorblockA{\IEEEauthorrefmark{1} Manhattan College, Riverdale, New York, USA.}\\
	\IEEEauthorblockA{\IEEEauthorrefmark{2} Western University, London, Ontario, Canada.}\\
	\IEEEauthorblockA{\IEEEauthorrefmark{3} Qatar University, Doha, Qatar.}}

\maketitle
\begin{abstract}
The rise in embedded and IoT device usage comes with an increase in LTE usage as well. About 70\% of an estimated 18 billion IoT devices will be using cellular LTE networks for efficient connections. This introduces several challenges such as security, latency, scalability, and quality of service, for which reason Edge Computing or Fog Computing has been introduced. The edge is capable of offloading resources to the edge to reduce workload at the cloud. Several security challenges come with Multi-access Edge Computing (MEC) such as location-based attacks, the man in the middle attacks, and sniffing. This paper proposes a Software-Defined Perimeter (SDP) framework to supplement MEC and provide added security. The SDP is capable of protecting the cloud from the edge by only authorizing authenticated users at the edge to access services in the cloud. The SDP is implemented within a Mobile Edge LTE network. Delay analysis of the implementation is performed, followed by a DoS attack to demonstrate the resilience of the proposed SDP. Further analyses such as CPU usage and Port Scanning were performed to verify the efficiency of the proposed SDP. This analysis is followed by concluding remarks with insight into the future of the SDP in MEC.
\end{abstract}


\section{Introduction}
\IEEEPARstart{T}{here} will be an estimated 18 billion IoT devices connected to the internet by 2022, 70\% of which will be on cellular LTE networks ( a 5\% share of LTE which is only expected to grow) \cite{erricson_mobility_report_iot}. Unfortunately, IoT systems in comparison to conventional computing systems have higher vulnerabilities and thus, more inherent security challenges. If these security challenges are not addressed properly, the wide adoption of IoT applications such as smart healthcare, smart cities, smart grids, smart transportation system, and smart agriculture cannot be realized \cite{IoTAPP}. For example, in a smart healthcare system, it is of utmost importance to secure sensitive information as well as critical assets in the system \cite{IoTSDP}. All IoT devices share common characteristics (such as low-cost design, high privacy requirement, high availability, and high trust-related issues) that makes securing IoT devices difficult with conventional security frameworks \cite{IoTChallenges}. For example, it was reported by Forbes.com that a successful attack that compromises a baby monitor device occurred in Houston \cite{hack}. Another incident was reported of a car been hijacked and stopped remotely by an attacker while the driver was driving on a highway \cite{hack2}. A survey conducted by CNN Money reveals that attackers have found vulnerability points of many IoT devices such as DVRs \cite{IoTDVR}, smart cameras \cite{IoTC}, and smart plugs \cite{IoTV} used in smart home settings. These reports necessitate the need to design a reliable security framework to secure IoT systems. Although the realization of IoT systems will make life easier and more convenient, it comes with several performance challenges such as security, latency, scalability, and quality of service. 

The concept of Edge computing has emerged as an innovative architecture that can improve the performance of IoT networks by acting as an intermediary to connect the edge devices to the cloud. For instance, edge computing can help reduce the response time and energy consumption of IoT networks. It can achieve this due to the presence of intermediary devices reducing traffic to the cloud servers. 

The European Telecommunications Standards Institute (ETSI) published the white paper on Multi-access Edge Computing (MEC) in September 2014, authored by the founders of the MEC industry initiative \cite{mec_whitepaper}.
In this paper, they discuss the applications and deployment scenarios for MEC. The pattern for each scenario generally consists of an LTE base station or evolved Node B base stations (eNB) through which all User Equipment (UEs) connect to the LTE network. The traffic is forwarded to the Evolved Packet Core (EPC) which consists of the Mobility Management Entity (MME), Home Subscriber Server (HSS), Serving Gateway (SGW), and Packet Data Network Gateway (PGW) from which traffic is redirected to the cloud. This pattern enables use cases such as activity device location tracking, low latency/ high bandwidth content delivery, edge-based video analytics, dynamic content optimization, and application-aware performance optimization \cite{mec_whitepaper}.

Despite the many advantages provided by adopting an edge computing-based architecture to connect edge and cloud computing (especially in terms of response time and energy consumption), several challenges/concerns arise in such architectures. Edge computing is an extension of the cloud and thus experiences similar security and privacy concerns. For instance, the inherent confidentiality issue worsened in the context of edge computing due to several applications (e.g., location-awareness service for mobile users) hosted at the edge of the network. This causes vulnerability to location-based attacks on end-devices where the user's information can be easily intercepted \cite{mec_whitepaper, survey}. Another security concern is the integrity of different entities (such as end-users, service providers, and infrastructure providers) in the edge computing ecosystem. This opens up trust and authentication issues between these entities that eventually will lead to attacks such as sniffing and man-in-the-middle attacks \cite{survey ,extended_fog}. End user's privacy (such as data and location) preservation is another critical issue that arises in the edge computing paradigm due to proximity to end devices. Sensitive information such as credit card numbers, personal emails, etc needs to be secured from attacks. Moreover, location-based services like the global positioning system (GPS) that require a user to share his/her location is another point of vulnerability \cite{survey}.

Edge computing requires a security framework to mitigate the aforementioned challenges. One framework capable of doing so is Software Defined Perimeter (SDP). The Cloud Security Alliance (CSA) proposed the SDP framework for securing any networking connected infrastructure. It does so by separating the data and control planes in a controller-based authentication model \cite{SDP}. Therefore, this paper proposes a combined MEC-SDP architecture that aims to secure LTE traffic at the edge before it reaches the cloud. The main contribution of this paper is summarized as follows:
\begin{itemize}
\item Propose a novel MEC-SDP architecture model that secures LTE traffic at the edge.
\item Provide an End-to-End delay analysis of the proposed architecture.
\item Verify the effectiveness of the proposed architecture against DOS and port scanning attacks.
\end{itemize}

The remainder of this paper is structured as follows: Section II discusses the security challenges and concerns facing the edge-to-cloud computing connection, as well as some of the related work proposed in the literature to tackle them.  In section III, the SDP framework as a potential solution is described in more detail in terms of its concept, architecture, and possible implementations. Moreover, the benefits gained by adopting it as a security framework within a network are presented. In section IV, some of the important attacks suffered in the mobile networking paradigm as reported in the literature are discussed. In Section V, a combined architecture is proposed and discussed in detail. In Section VI, the MEC-SDP combined architecture is implemented and delay analysis of the testbed is presented. In section VII, the discussion of test results is presented. Lastly, in Section VIII,  the paper is concluded by reiterating the importance of the presented study and discussing the potential for future work.

\section{Related Work}

\subsection{Edge Computing Security}

\begin{figure}
\centering
\includegraphics[width=0.5\textwidth]{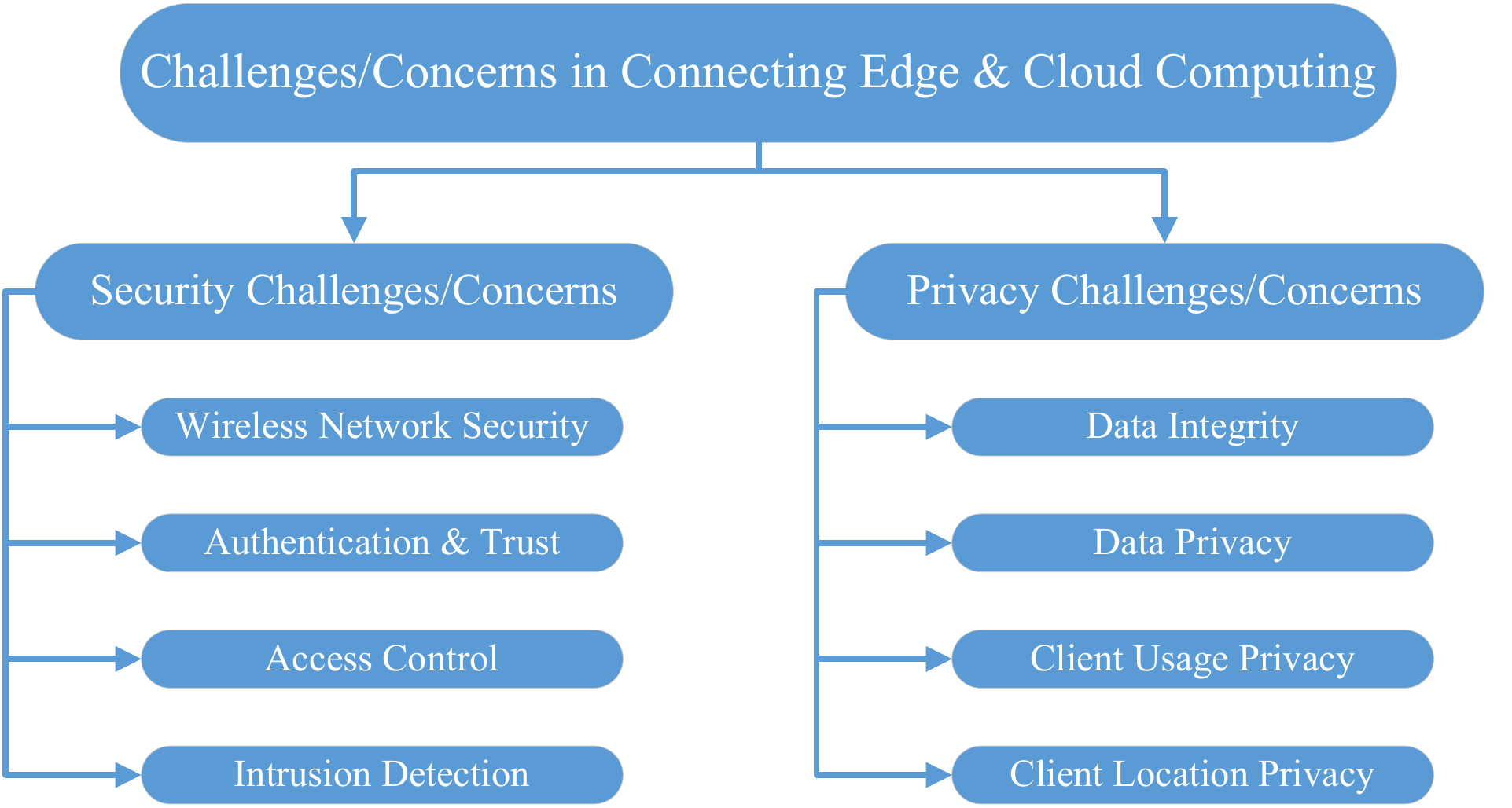}
\caption{Challenges}
\label{fig:challenges}
\end{figure}
 Edge computing is an emergent technology where cloud computing-like capabilities are extended to the edge of the network. This significantly reduces the latency experienced in cloud computing framework and provides seamless integration with various application service providers and vendors \cite{MEC}. However, adopting such architecture introduces numerous challenges/concerns.  As shown in Fig. \ref{fig:challenges}, these challenges/concerns can be divided into two main categories: security challenges/concerns and privacy challenges/concerns. In what follows, a brief description of each of these challenges is given along with some of the related work done in the literature to address them.
 
 The security category comprises of wireless network security challenges, authentication and trust challenges, access control, and intrusion detection. Edge Computing takes advantage of several different technologies to build the network introducing potential to several attacks such as man-in-the-middle-attack, DoS attacks, DDOS attack, and wireless jamming \cite{extended_fog}. The use of virtualization technology within the edge computing platform also introduces potential security threats such as virtual machine (VM) hopping or eavesdropping.  
 
 Several studies in the literature have focused on mitigating the effect of some of the aforementioned challenges in the context of edge computing. For example, the authors in \cite{DDOS} proposed a per-packet based detection mechanism that integrates packet filtering techniques with a congestion control framework to mitigate DDOS attacks. Whenever a malicious packet is identified, it will be dropped before reaching its target. Another study \cite{EDDOS} proposed a mechanism that exploits the fact that packets from the same path have a similar identifier to detect DDOS attacks. By this analogy, packets that have the same identifier as previously detected DDOS-oriented packets are most likely to be a DDOS attack. In \cite{UDP}, the authors proposed a negative selection algorithm that aims to detect the legitimacy of a user based on a set of eigenvalues to resist a DDOS attack. All the aforementioned mechanism requires per-packet information such as packet identifier and IP/MAC addresses to effectively detect DDOS attacks which an attacker can manipulate. This clearly illustrates the necessity of adopting a stronger solution that can prevent DOS and DDOS attacks without relying on any packet information. 
 
 There is a significant amount of literature that demonstrates the advantage of adopting machine learning and deep learning techniques in detecting DDOS attacks \cite{MLDDOS,ETDDOS,DLDDOS}. In \cite{MLDDOS}, the authors utilize some basic machine learning techniques such as Bayes and Bayesian network classifiers to successfully detect botnet DDOS attacks. A deep learning model was adopted in \cite{DLDDOS} to detect encrypted DDOS traffic using a simple auto-encoding mechanism. Similarly, the authors in \cite{EDDOS} use neural networks to detect DDOS attacks. These learning-based detection approaches require a significant amount of DDOS traffic for learning/training purposes, which can only be acquired after the edge servers are exposed to the attacks. Therefore, a better security framework is much needed to fully adopt an edge computing platform.
 
 Another security threat that sees a lot of research effort is authentication and authorization between edge devices and edge servers. The author in \cite{AUTH1} proposed adopting an active jammer with a wireless injection mechanism to mitigate the brute-force attack that occurs while decrypting WPA traffic. In \cite{AUTH2}, a revised version of the original key exchange process in WPA/WPA2 protocol is proposed to reduce the vulnerabilities in authentication schemes. A black-box verification approach is adopted in \cite{AUTH3} to tackle impersonation attacks that weaken authentication procedures.
 
 From the aforementioned discussion, it is evident that a resilient security framework that can mitigate various attacks simultaneously is needed. SDP is one such framework capable of meeting a network's security requirements.

\begin{figure}[h]
    \centering
    \includegraphics[width=0.5\textwidth,]{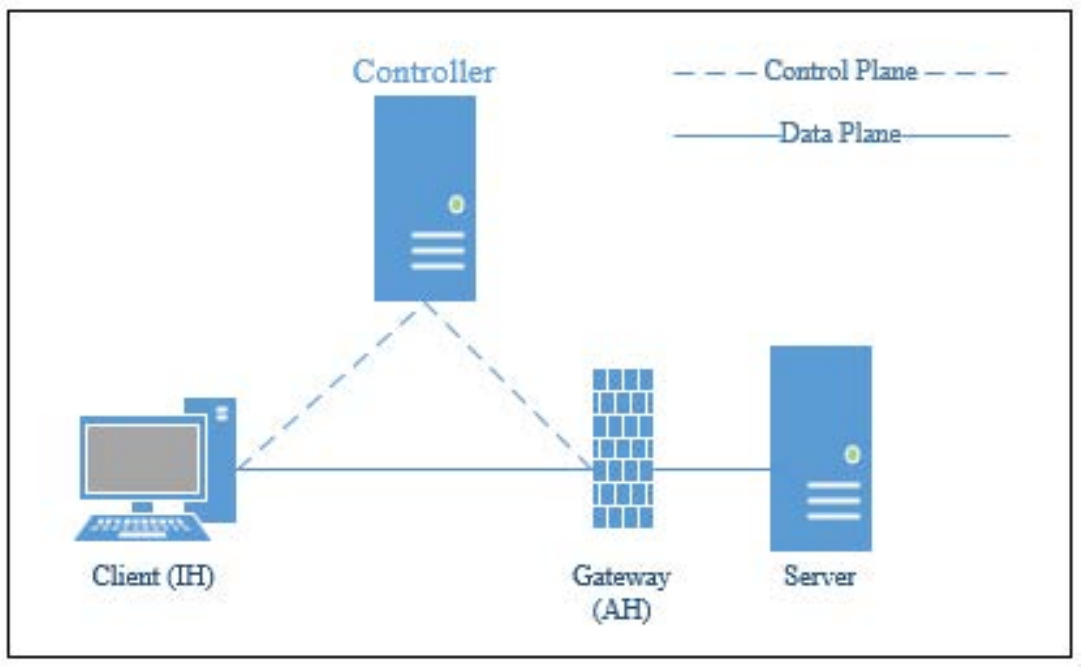}
    \caption{SDP Architecture}
    \label{fig:sdppic}
\end{figure}

\begin{figure*}[ht!]
\centering
\includegraphics[width=1\textwidth, height=8.7cm]{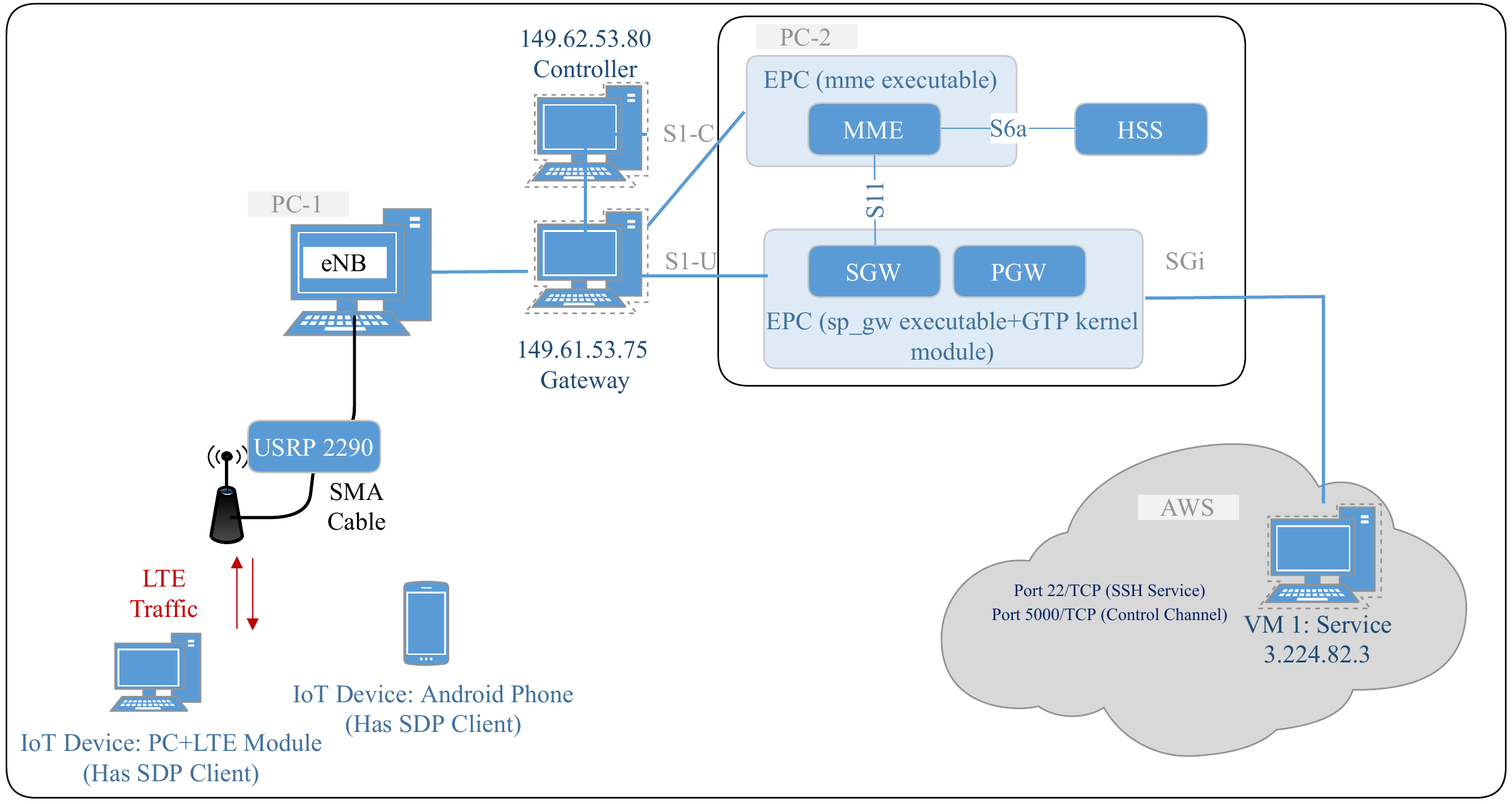}
\caption{MEC System Protected with SDP}
\label{fig:System}
\end{figure*}
\subsection{Software Defined Perimeter (SDP)}

SDP is a zero-trust security framework, meaning it works on a need to know authentication model for every service regardless of where in the network you accessing it from \cite{SDP_CSA2}. This pattern is versatile for many applications and as such, SDP has been explored in several applications and thoroughly tested. It has recently been proposed as a security framework for both Network Function Virtualization and Software Defined Networking. Both suffer from challenges such as hypervisor hijacking, DoS attack vulnerability, Virtual Machine hopping, and backdoor attacks. With the introduction of SDP, security threats were successfully mitigated without sacrificing performance \cite{8826550, nfv_sdp}. SDP has also been explored for IoT purposes; the traditional authentication methods used in Message Queuing Telemetry Transport is replaced with SDP's single packet authorization mechanism. This study demonstrated SDP's capability to introduce security and privacy to vulnerable and low bandwidth IoT networks \cite{SDPIOT}.  

This paper differs from previous studies in that it explores the advantage of introducing SDP to MEC. The SDP components are split amongst the edge and the cloud in this architecture and combined with the LTE core network for the first time.

\section{Software Defined Perimeter Framework}
\subsection{SDP Overview}
\begin{figure*}[ht!]
\centering
\includegraphics[width=1\textwidth, height=9cm]{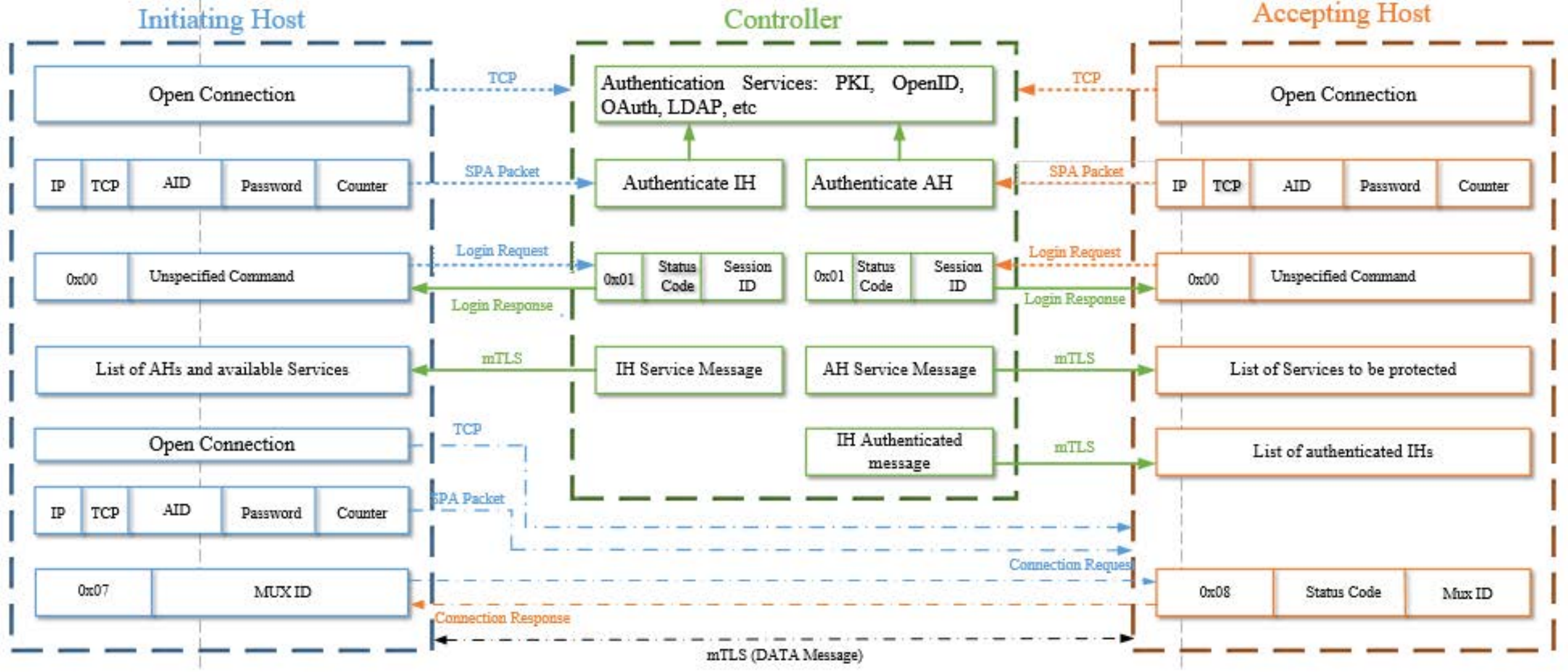}
\caption{Conceptual Architecture for SDP Workflow}
\label{fig:CSDP}
\end{figure*}

Due to the ever-increasing network security challenges brought by the rapid development of today's networking paradigm (especially the 'software-ization' of network resources), the traditional security model that aims to secure network resources within a defense perimeter can no longer suffice network security needs. Hence, many organizations have proposed security specifications for a robust software-based security framework that meet the current network security needs.
 \newline
 \indent The National Institute of Standards and Technology (NIST) defined a zero-trust architecture (ZTA) to plan an enterprise's infrastructure and workflows that follow a zero trust model as the basis for its architecture \cite{ZTA}. SDP, another security framework that adopts the zero trust model was proposed by the CSA. Table 1 presents a comparative analysis of the two frameworks in terms of access authorization responsibility, framework components, and practical implementation. Considering the promising results in mitigating numerous attacks obtained by adopting SDP \cite{SDP,nfv_sdp,8826550}, and the  open source implementation of SDP components by Waverley Labs, we chose to adopt SDP in this work.
  \newline
 \indent Typically, the SDP framework consists of three actors; SDP controller, SDP initiating host (IH), and SDP accepting host (AH). Using a variety of combinations of these components, SDP can be implemented to suit multiple frameworks such as client-to-gateway, client-to-server, server-to-server, and client-to-server-to-client \cite{SDP_CSA}. For the purposes of this work, we adopt the client-to-gateway architecture as shown in Figure~\ref{fig:sdppic} because it provides seamless integration with the proposed architecture. Under this framework, the service(s) is/are protected behind the AH module such that the AH module acts as a gateway between the clients and the protected service(s). Similar to Software-Defined networking, SDP decouples the control plane and the data plane. The SDP controller resides in the control plane and serves as the brain that makes the final decision on the legitimacy of any client before granting access to any service. Figure~\ref{fig:CSDP} displays a conceptual architecture for SDP workflow. 
  \newline
 \indent When an SDP controller is brought online, it connects to the appropriate authentication and authorization services such as Public Key Infrastructure (PKI) issuing certificate authority services, OpenID, SAML, OAuth, LDA, etc. All hosts (i.e., either initiating or accepting) in the SDP network must authenticate to the controller by sending an SPA packet as shown in Figure~\ref{fig:CSDP}. For interested readers regarding the details of packets' format (i.e., SPA packet, Login request, Login response, connection request and connection response) depicted in Figure~\ref{fig:CSDP}, refer to \cite{SDP_CSA1}. The SDP controller keeps a list of authorized hosts and services in a MySQL database and generates keys and certificates for each host in the database. After a client is successfully authenticated, the controller sends an IH services message containing a list of available AHs and services to the client. Then the SDP controller sends a message to the gateway to configure its firewall rules to allow the client's request to pass and access a service through the AH services message and IH authenticated message. After that, the gateway establishes a Mutual Transport Layer Security (mTLS) connection between the client and itself for a defined period for data transfer. It is worth mentioning that when this defined period ends, the rule to accept the client's request will be deleted but the connection is left open and monitor by the tracking mechanism within the AH module.
 
\subsection{SDP Security Benefits}
Any security framework must guarantee a certain level of certainty in terms of resource protection, data integrity, access control, availability issues, and confidentiality related issues. As a matter of fact, the SDP framework combines five layers of security to tackle these uncertainties. These security layers are single packet authentication (SPA), mutual transport layer security (mTLS), device validation (VD), dynamic firewall, and application binding (AppB) \cite{SDP}. The SPA packet is the first message sent to initiate connection paths either between the controller and accepting host or initiating host and accepting host. Then a secure mTLS connection is established only when the SPA packet is valid and verified by the controller. Despite having been authenticated, the device validation process is used to validate the integrity of users to ensure the credential keys are not stolen by a third party intruder. A dynamic firewall policy ensures all traffic is blocked at all times except when authorized by the controller and application binding process forces users to only communicate through the secure mTLS connection. The combination of these five protocols provides a strong security framework capable of mitigating many network-based attacks such as DOS attack, MITM attack, port scanning attack, and so on.
\newline
 \indent Looking at confidentiality related issues such as leakage of sensitive private information or credential keys, the SDP framework not only provides a secure mTLS connection to ensure no data is retrieved by the unauthorized users but also has a device validation process to track the legitimacy of users throughout the user's session. 
\newline
 \indent In the context of security, the availability of components that make up the security framework is of utmost importance to ensure always-on protection of network resources. Failure of any components might give attackers a window to breach the network. Due to the 'software-ized' nature of SDP framework, the controller and gateway are readily available as software installed within the network. The effect of software crashes is minimal as the time re-instantiate the SDP components is short \cite{8826550}.

\setlength{\arrayrulewidth}{0.5mm}
\setlength{\tabcolsep}{10pt}
\renewcommand{\arraystretch}{1.5}
\begin{table*}[ht!]
\begin{center}
\caption{Software defined perimeter versus zero trust architecture}
\begin{tabular}{ | m{6em} | m{21em}| m{21em} |  }
  \hline
 Features & Software defined perimeter (SDP) & Zero trust architecture (ZTA) \\
  \hline
 Main purpose & Securing defined network resources from illegitimate clients using a dynamic firewall mechanism with drop-all policy for incoming traffic \cite{SDP_CSA1}. &  Securing enterprise resources from unauthorized users through authentication and authorization procedures to reduce the implicit trust zones within the enterprise's network \cite{ZTA}. \\
  \hline
 Adopted Model & A zero trust security model (i.e., no client is trusted even if it resides within the network) \cite{SDP_CSA2}  & A zero trust security model \cite{ZTA}  \\
  \hline
 Components  & \begin{itemize}
\item SDP controller: This component is responsible for creating credential keys and authentication of clients.\cite{SDP_CSA1}
\item SDP accepting host: This component is responsible for blocking all unauthorized traffic and works closely with the controller to enforce these rules.\cite{SDP_CSA1}
\item SDP initiating host: This component is the legitimate client stored in the controller database and given a credential keys for authentication.\cite{SDP_CSA1}
\end{itemize} &  \begin{itemize}
\item Policy Engine (PE): This component is the main brain responsible for final decision to grant legitimate users access to any network resource. \cite{ZTA} 
\item Policy Administrator: This component works closely with the PE to create credential keys and authentication tokens that will be use to grant access to legitimate users. Moreover, it is responsible for establishing connection path between legitimate users and authorized resource. \cite{ZTA}
\item Policy Enforcement Point: The responsibility of PEP is to enable, monitor and eventually terminate connection path between legitimate users and authorized resources. \cite{ZTA}
\end{itemize} \\

 \hline
 Access authorization & The SDP controller makes the decision to grant or deny access to network resources using the generated credential to authenticate and verify client requests \cite{SDP_CSA1,SDP_CSA2}. & The PE uses the enterprise defined policies and a set of external information (such as ID management system, enterprises' public key infrastructure (PKI), data access policies, threat intelligence feeds and continuous diagnostic and mitigation systems) as an input to a trust algorithm to grant or deny access to network resources. \cite{ZTA} \\
 \hline
Implementations & Waverly Labs OpenSDP  Project&  NA\\
 \hline 
\end{tabular}
\end{center}
\label{table:SDPZTA}
\end{table*}

\section{Attacks in mobile networking}
Mobile networking suffers from several potential attack vectors introduced by a shift towards an IP-based architecture and new radio access technologies. These attacks can compromise confidentiality, integrity, or availability of mobile networks \cite{survey_mobile_attacks}. In \cite{survey_mobile_attacks}, several attack types are explored including security and confidentiality attacks, IP based attacks, signaling attacks, and jamming attacks. The attack targets for these include DoS, Voice over LTE (VoLTE), General Packet Radio Service Tunneling Protocol (GTP), physical layer, and LTE security. Attacks on VoLTE include message forgery, flooding, and message tampering which may result in VoLTE degradation, over-billing, voice phishing, and even DoS \cite{survey_97}. GTP can be taken advantage of to compromise information, exhaust network resources, perform phishing, and interrupt services \cite{6975521}.  Denial of service attacks can occur in a variety of ways including flooding, redirection, interference, and spoofing. It is an example of an attack target which falls under each of the aforementioned attack types. The attacks such as VoLTE and GTP are highly specific to just specific features of LTE. Voice over LTE is an optional feature offered by LTE networks to place voice calls over LTE. GTP utilized SIP-based messaging that has security risks which can be mitigated by strong SIP message checking. DOS attacks on the contrary are highly generic and can result in Loss of Confidentiality, Availability, Integrity, Control or even Theft of service.       \newline
\indent The choice of network-based attack for this paper is the DoS attack. According to Nexusguard, the year on year increase of this type of attack is up 85.66\% in 2019 Q3 \cite{nexusguard}. There are several types of DoS attacks such as low-rate DoS which use short period bursts of traffic to keep the router buffer full \cite{low_rate_DOS} \cite{low_rate_DOS2}, HTTP flood attack which sends an influx of valid GET requests to an HTTP server incapacitating it \cite{http_dos}, or TCP SYN Flood attacks which used spoofed IPs to take advantage of the TCP three-way handshake \cite{tcp_syn}. The attack in this paper is a TCP SYN Flood attack which has seen a 177.37\% year-on-year growth. This attack, as previously discussed,  targets vulnerabilities in the widely adopted three-way handshake which makes is such a versatile threat.

\section{Proposed Solution}

The combined MEC-SDP architecture in this paper is being proposed for the first time. MEC allows for reduced energy consumption, improved response times, and introduces several new use cases but comes with some security disadvantages as mentioned prior. SDP provides an added layer of security by separating the control and data planes and using SPA based authentication, enabling us to secure LTE traffic to the cloud at the Edge.

The proposed architecture, depicted in Figure~\ref{fig:System} consists of a Universal Software Radio Peripheral (USRP) based eNB setup. The USRP can transmit and receive an LTE signal as well as connect to a host machine running the eNB over a high-speed link. LTE devices with configured SIM cards can connect to the USRP and as a result, be connected to the EPC. The EPC consists of HSS, MME, SGW, and PGW. The HSS maintains user subscription information for UEs. The MME handles the control plane by interacting with the HSS. The SGW and PGW are the inbound and outbound gateways, respectively. The SGW acts as a router between the eNB and the PGW and the PGW uses the SGi interface to communicate with the edge gateway. This ultimately provides the UEs with a route to the public internet via the gateway. An SDP controller and gateway are situated on their own Linux 16.04 machines which are used to secure the cloud services from UEs at the edge. A VM in the cloud is incorporated to allow us to connect to the ssh service. The user keys and certificates are generated by the controller and distributed, for the purposes of this paper that distribution is a manual process.

\begin{figure*}[b]
\centering
\includegraphics[width=1\textwidth]{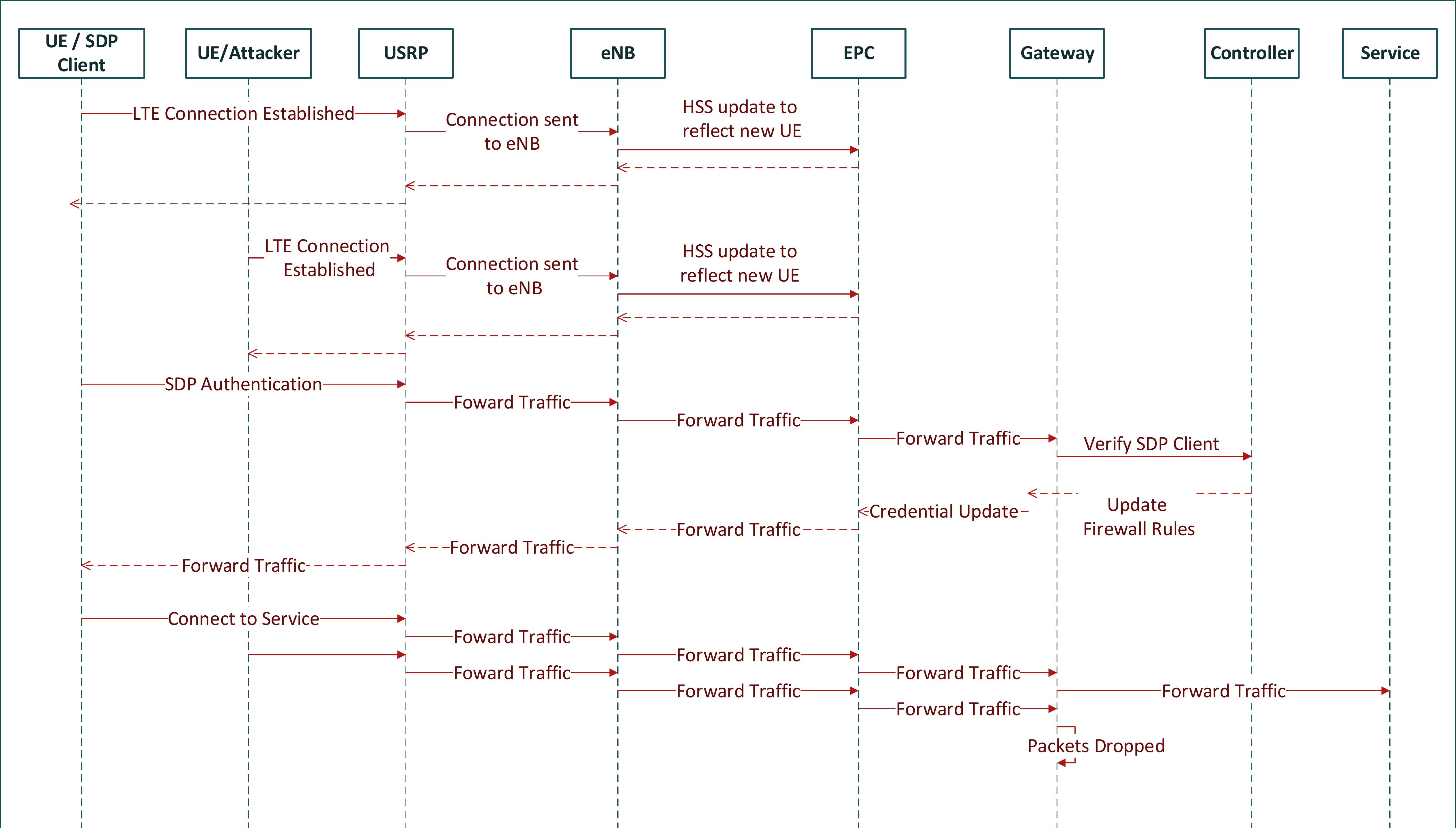}
\caption{Sequence Diagram}
\label{fig:sequence}
\end{figure*}

UEs are configured with the SDP client and connected to the USRP and eNB using Commercial-Off-The-Shelf (COTS) LTE dongles. The eNB forwards traffic to the EPC, which then forwards all traffic to the SDP gateway machine. To access the service which has been deployed in the cloud, the UE must authenticate with the controller at the edge first. Without authenticating with SDP, no internet access is granted at all. The end to end interaction with the system is depicted in Figure~\ref{fig:sequence}. Algorithm 1 provides a detailed authentication process for the proposed MEC-SDP architecture. When a UE with a SIM configured for the LTE network joins the network, it establishes a connection via the USRP and eNB. The eNB relays this to the EPC, in which the HSS updates to account for the new UE and allows the UE to access the internet. In order for the UE to access the service in the cloud, however, it must be authorized using SDP. The UE client sends the initial SPA message to the gateway at the edge which communicates to the controller. If authorized, all connected gateways update their firewall via FWKNOP to allow only the access level specified by the SDP controller. The UE should now be able to connect to via the gateway to the services they have been authorized to, which is only the service.

This edge controller approach offloads the control plane to the edge of the network, improving response times for authentication to the SDP network. SDP can also be used to secure edge services that prevent unauthorized access to services such as location monitoring, mitigating the potential for location-based attacks. Also, by moving the controller to the edge with its gateway, the controller is protected from targeted attacks from unsecured networks as well.

\section{Test-bed and Delay Analysis}

\begin{figure*}[h!]
    \centering
    
        \includegraphics[width=1\textwidth]{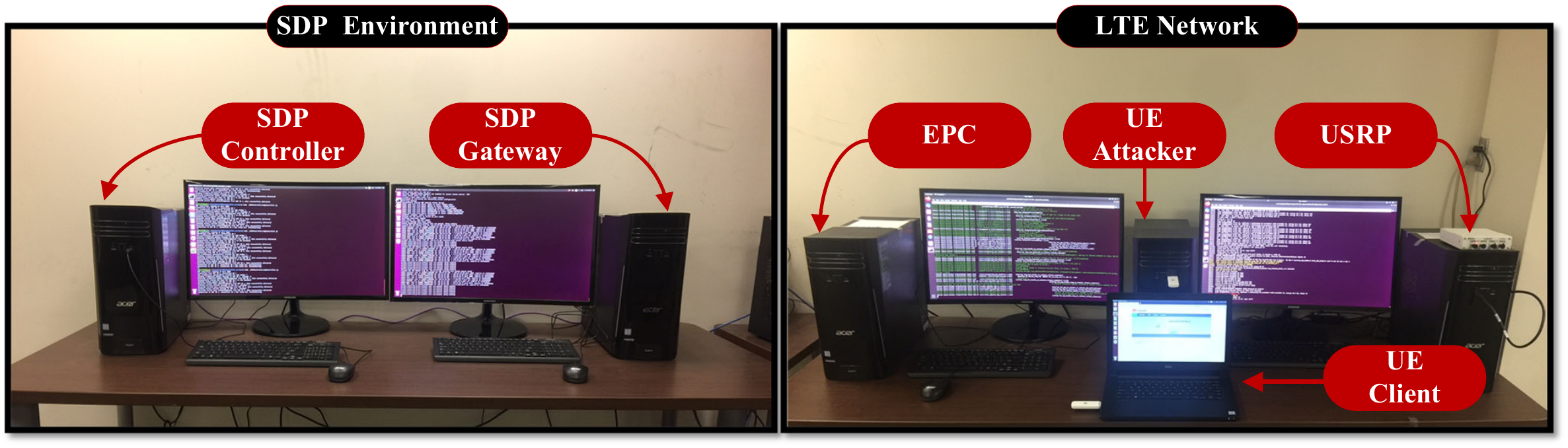}

        \label{fig:sdp}
  \caption{Testbed Setup}\label{fig:testbed}
\end{figure*}
The testbed used in this paper consists of two open-source tools. For the LTE Core and Radio Access Network, OpenAirInterface (OAI) open-source project and for the SDP network, Waverley Labs' Open SDP project. The system is comprised of 6 physical machines and an AWS EC2 instance. Figure~\ref{fig:testbed} displays the actual machines emulating the complete LTE network and those emulating the SDP environment. Descriptions of each component in the testbed is provided in Table 2.

\vspace*{-.35cm}
\begin{algorithm}[h]

  The LTE network is brought online\;
  
  Assume the edge GTWY (AH) is already initialized and authorized by the edge CTRL\;
  
  The UE client sends an SPA packet to the edge controller\;
  
  \eIf{the SPA packet is valid}{
   The edge CTRL verifies the UE client (using the stored certificate) and establishes an mTLS secure connection between itself and the UE client\;
   
   The edge CTRL sends all information about the UE client’s authorized service (s) to the edge GTWY\;
   
   The UE client sends SPA packet to the edge GTWY\;
   
   \eIf{the SPA packet is valid}{
   The edge GTWY verifies the UE client using the certificate provided by the edge CTRL\;
   
   The edge GTWY sets up corresponding firewall rules to allow UE client access to the authorized service (s) for a defined period, T\;
   
   \eIf{T $>$ 0}{
   The UE client attempt to ssh into the AWS instance protected by the edge GTWY\;
   
   \While{connection tracking is still valid }{
  Send test packets\;
    }
   }{
   The edge GTWY removes corresponding firewall rules 
   }
   }{
   Block all connection request from the UE client\;  
   }
   }{
  Drop all packets
  }
 \caption{UE client authentication process for the proposed MEC-SDP framework}
\end{algorithm}

\begin{table*}[h]
\begin{center}
\caption{Description of the Testbed}
\begin{tabular}{ | m{10em} | m{40em}| }
  \hline
 Machines & Description \\
  \hline
 UE + Client & To emulate the legitimate UE, an LTE dongle (Huawei LTE USB stick, E3372) using an an open-cells programmable USIM card is configured for our LTE network. This LTE dongle is then connected to a Linux machine (Ubuntu 16.04) that has SDP client module offered by Waverley Labs. This setup allows us to connect to the USPR device emulating eNodeB and also connects to the SDP network as a legitimate user \\
  \hline
 USRP + eNodeB module & The USRP is the National Instruments (NI) USRP-2901 model which is connected to a Linux machine (ubuntu 18.04) running the OpenAirInterface5g eNodeB configuration over high speed link  \\ 
 \hline
 EPC module & The EPC module is running on Linux machine (Ubuntu 18.04). We adopt the latest developed version of the OpenAirInterface EPC module for this work \\ 
 \hline
 UE + Attacker & To emulate the illegitimate UE acting as the attacking entity sending a large volume of requests to our service as well as performing port scanning attack, an LTE dongle (Huawei LTE USB stick, E3372) using an an open-cells programmable USIM card is configured to our LTE network. This LTE dongle is then connected to a Linux machine (Ubuntu 16.04) which allows us to connect to the USPR device emulating eNodeB. The attack is performed using the HPING3 testing suite \\ 
 \hline
 SDP Controller & The SDP controller provided with the control module is running on a Linux machine (ubuntu 16.04). All UEs must authenticate with the SDP controller before gaining access to the service on the cloud \\ 
 \hline
 SDP Gateway & The SDP gateway is running on Linux machine (Ubuntu 16.04). It is configured with a drop-all policy for all traffic except from the controller \\ 
 \hline
 AWS Cloud Instance & In AWS, there is an EC2 instance running Linux Ubuntu 16.04. The VM is used for a basic SSH service which the gateway is protecting from unauthorized UEs \\ 
 \hline
\end{tabular}
\end{center}
\label{table:DE}
\end{table*}

\vspace*{-.4cm}

\subsection{Test Case}

Firstly, the LTE network is brought online starting with the EPC module, then the eNodeB module, and finally the UE. For the EPC module, the four entities are launched sequentially starting with the HSS entity then the MME entity followed by the SPGW-U entity and finally the SPGW-C entity. After the EPC module stabilizes, the eNodeB is brought online which automatically connects to the MME entity. Lastly, the UE is brought online and is automatically attached and connected to the EPC via eNodeB.

Then the connection is established on the authorized client UE. To do this, the LTE network is joined and then the SDP client software is run. This will send an SPA packet to the SDP gateway to authenticate. The gateway verifies with the controller and will update its firewall accordingly to allow communication from the client if it has permissions. The client can now SSH to the cloud service using the NAT gateway. The configuration is such that an SSH session on port 4444 of the gateway will redirect and consequently establish an SSH session with the service in the cloud. The attacker machine also established a connection to the same LTE network but does not authenticate to the controller before attempting to flood the cloud service. 

To evaluate the performance of the proposed framework, the theoretical delay is compared with the actual delay incurred in the implemented testbed. This delay analysis includes the end-to-end delay (i.e., the time taken by a UE client to connect to the ssh service on the cloud) with and without SDP, the controller connection overhead, and the gateway connection overhead. The gateway connection is then monitored for 120s, during which a SYN flood attack is launched for 60 seconds. The traffic is all captured at the SDP gateway and then analyzed. This is finally followed by a port scan attack to verify the gateway configuration.

\subsection{Delay Analysis}

Any packet traversing through a computer network is subjected to various types of delay along the path from source to the destination node. The most important among these delays are transmission delay, propagation delay, processing delay and queuing delay. In this work, we assume the processing, and queuing delay is negligible relative to the other two delays. Thus the total delay from one node to the next is the summation of the transmission and propagation delay. This can be represented mathematically as follows:
\begin{equation}
Total \quad Delay = \frac{\alpha}{R} \quad + \quad \frac{\beta}{S}
\end{equation}

\begin{figure}[h]
    \centering
    \includegraphics[width=.5\textwidth]{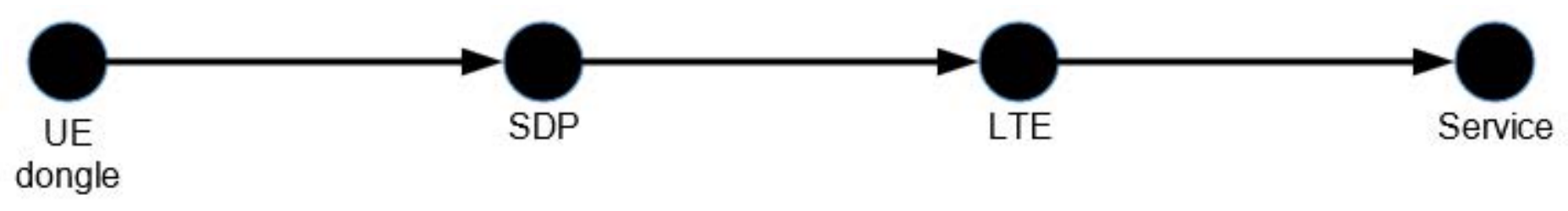}
    \caption{Nodal representation}
    \label{fig:node}
\end{figure}

Where $\alpha$ denotes the size of the packet, $R$ denotes the transmission rate of the link joining the two nodes under analysis, $\beta$ denotes the distance between the two nodes and $S$ is the propagation speed of the link. The proposed testbed has 4 nodes as shown in Figure~\ref{fig:node}. 

\begin{itemize}
\item Case 1: Delay in the Initialization phase
\end{itemize}

\begin{figure}[h]
    \centering
    \includegraphics[width=.5\textwidth]{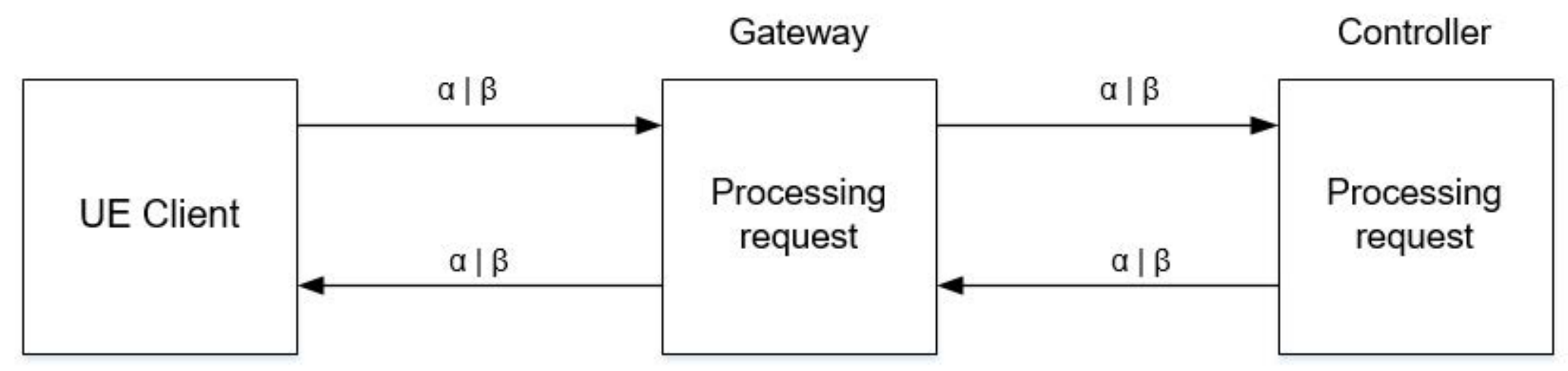}
    \caption{Controller and gateway overhead}
    \label{fig:clgc}
\end{figure}

The delay incurred in the initialization phase involves the time it takes to initiates the SDP and the LTE network. In particular, we consider the overhead introduced for verification and authentication process by the SDP controller and the gateway as well as the connection setup time to the LTE network. Figure~\ref{fig:clgc} shows the exchange of packets that occurs between the client and the controller for verification and authentication purposes. The gateway relays all the packets sent by the client to the controller and vice versa. Thus, the overall SDP overhead  is calculated as follows:

\begin{multline}
SDP \quad overhead = 3[\frac{\alpha_{1}}{R} + \frac{\beta_{1}}{S}] + 4[\frac{\alpha_{2}}{R} + \frac{\beta_{2}}{S}]\\
+ 3[\frac{\alpha_{3}}{R} + \frac{\beta_{3}}{S}] + 5[\frac{\alpha_{4}}{R} + \frac{\beta_{4}}{S}]
\end{multline}

Where $\alpha_{1}$ and $\beta_{1}$ denote the size of the packet sent from the UE Client to the gateway and the length of the link joining them respectively. Likewise $\alpha_{2}$ and $\beta_{2}$ denote the size of the packet sent from the gateway to the controller and the length of the link joining them respectively. $\alpha_{3}$ and $\beta_{3}$ denote the size of the packet sent from the controller to the gateway and the length of the link joining them respectively and lastly $\alpha_{4}$ and $\beta_{4}$ denote the size of the packet sent from the the gateway back to the UE Client and the length of the link joining them respectively. 

For the LTE network, we consider the delay suffered for connecting the UE to the eNB and the EPC module and is calculated as follows:

\begin{equation}
LTE \quad Delay = \frac{\alpha_{5}}{R} + \frac{\beta_{5}}{S} + \frac{\alpha_{6}}{R} + \frac{\beta_{6}}{S}
\end{equation}

Where $\alpha_{5}$ and $\beta_{5}$ denote the size of the packet sent from the UE Client to the eNB and the length of the link joining them respectively and $\alpha_{6}$ and $\beta_{6}$ denote the size of the packet sent from the eNB to the HSS module and the length of the link joining them respectively.

Therefore the delay suffered in the initialization phase of the proposed testbed is calculated as follows:

\begin{multline}
Delay = [\frac{1}{R} + \frac{1}{S}][3\alpha_{1} + 4\alpha_{2} + 3\alpha_{3} + 5\alpha_{4} + \alpha_{5} + \alpha_{6}\\ 
+ 3\beta_{1} + 4\beta_{2} + 3\beta_{3} + 5\beta_{4} + \beta_{5} + \beta_{6}]
\end{multline}

\begin{itemize}
\item Case 2: End-to-End Delay of the proposed testbed
\end{itemize}

The End-to-End delay of this work comprises of the initialization phase delay and the delay suffered to ssh into the AWS service as described in the previous section. It's worth mentioning that the controller overhead occurs only once for any legitimate client. After a client is successfully registered and verified by the controller, the only overhead in any other session is the gateway overhead for updating the necessary firewall rules to allow access to authorized service. Consequently the End-to-End delay with SDP is calculated as follows:

\begin{multline}
End-to-End \quad Delay = [\frac{1}{R} + \frac{1}{S}][3\alpha_{1} + 4\alpha_{2} + 3\alpha_{3} + 5\alpha_{4} + \alpha_{5}\\
+ \alpha_{6} + \alpha_{7} + \alpha_{8}
+ 3\beta_{1} + 4\beta_{2} + 3\beta_{3}\\
+ 5\beta_{4} + \beta_{5} + \beta_{6} + \beta_{7} + \beta_{8}]
\end{multline}

Where $\alpha_{7}$ and $\beta_{7}$ denote the size of the packet sent from the UE Client to the cloud service and the length of the link joining them respectively and $\alpha_{8}$ and $\beta_{8}$ denote the size of the packet sent from the cloud service to the UE client and the length of the link joining them respectively.

\section{Results and Discussion}

As expected the end-to-delay with SDP is a bit higher than without SDP as shown in Table 3. This is because the UE client not only needs to authenticate with the SDP controller but has to wait for the gateway to update the firewalls before the service can be accessed. Next, the controller overhead as well as the gateway overhead is determined by analyzing the exchange of packets between the UE, gateway and the controller using Wireshark tool. The process was monitored 10 times and the average is recorded as 0.0477037 seconds and 0.04892312 seconds respectively as depicted in Table 3. This overhead is very much tolerable compared to the security benefits gained by adopting SDP framework. Additionally, the calculated theoretical values in the End-to-End delay with and without SDP were very much closed to the actual measured values from the implemented testbed which verifies the testbed correctness.

\begin{table}[h!]
\begin{center}
\caption{Delay Analysis Evaluation}
\begin{tabular}{ | m{8em} | m{5em}| m{5em}| }
  \hline
 Delays (sec) & Theoretical & Measured\\
  \hline
 End-to-End delay with SDP & 0.397653420 & 0.555149629 \\
  \hline
 End-to-End delay without SDP & 0.278936738 & 0.547851958 \\ 
 \hline
 Controller overhead & ~NA &  0.04477037\\ 
 \hline
 Gateway overhead & ~NA &  0.04892312 \\ 
 \hline
\end{tabular}
\end{center}
\label{table:PE}
\end{table}

\begin{figure}[h!]
    \centering
    \includegraphics[width=.5\textwidth]{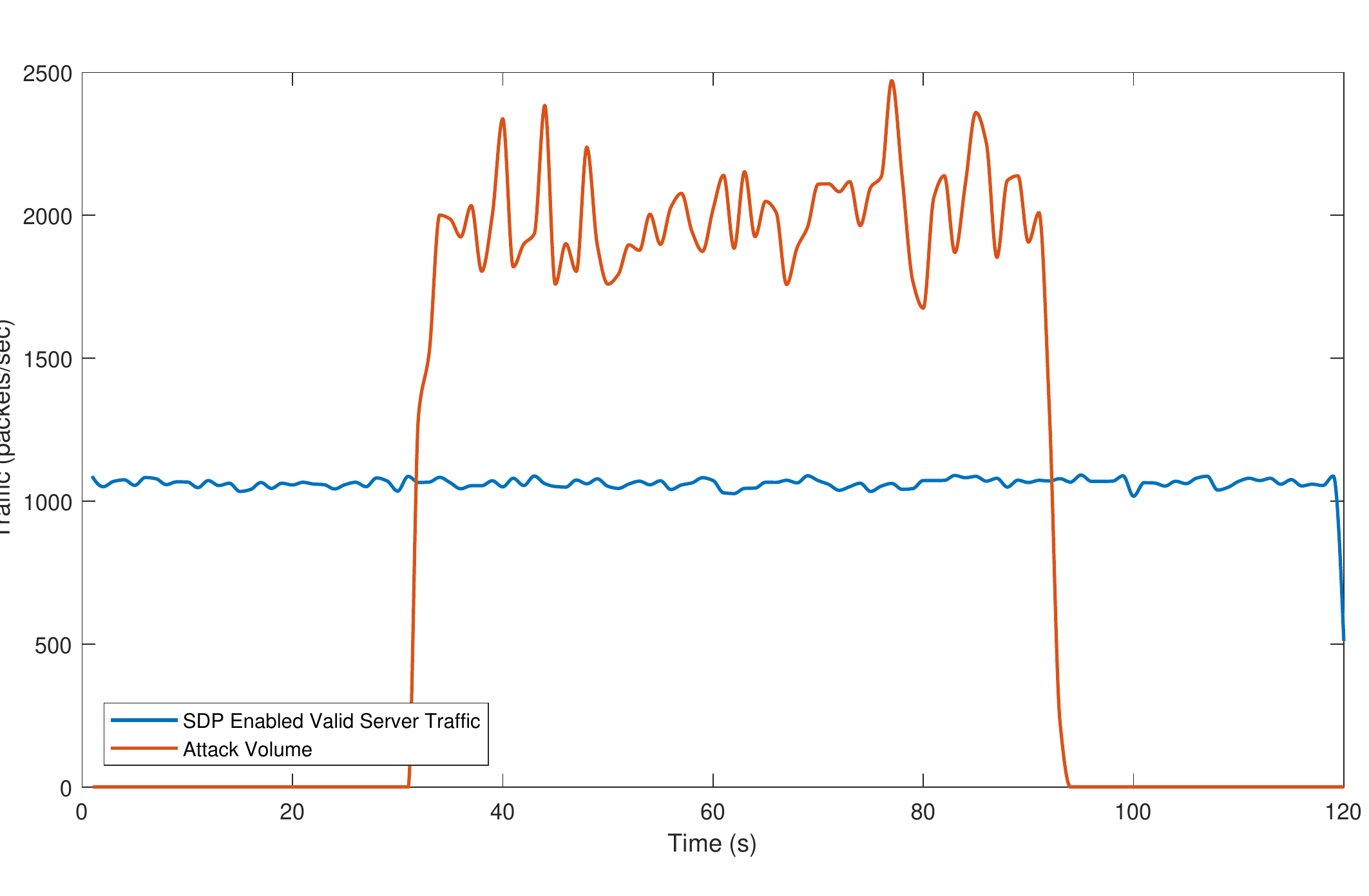}
    \caption{DOS attack capture}
    \label{fig:attack}
\end{figure}
The DOS attack trails were performed 3 times for 2 minutes each trail and the average of the captures is shown in Figure~\ref{fig:attack}. The blue curve is the measure of traffic with an ACK flag, which is 1-to-1 relation with packet volume being sent between the service and the authorized client. The orange curve is a measure of attack packets, which can be distinguished by the frame length of fewer than 60 bytes, as DoS attacks attempt to send a high volume of very small-sized packets. None of the attack packets were acknowledged as they are dropped by the gateway. The graph also shows that performance did not suffer during the duration of the attack. 

\begin{figure}[h]
    \centering
    \includegraphics[width=0.5\textwidth]{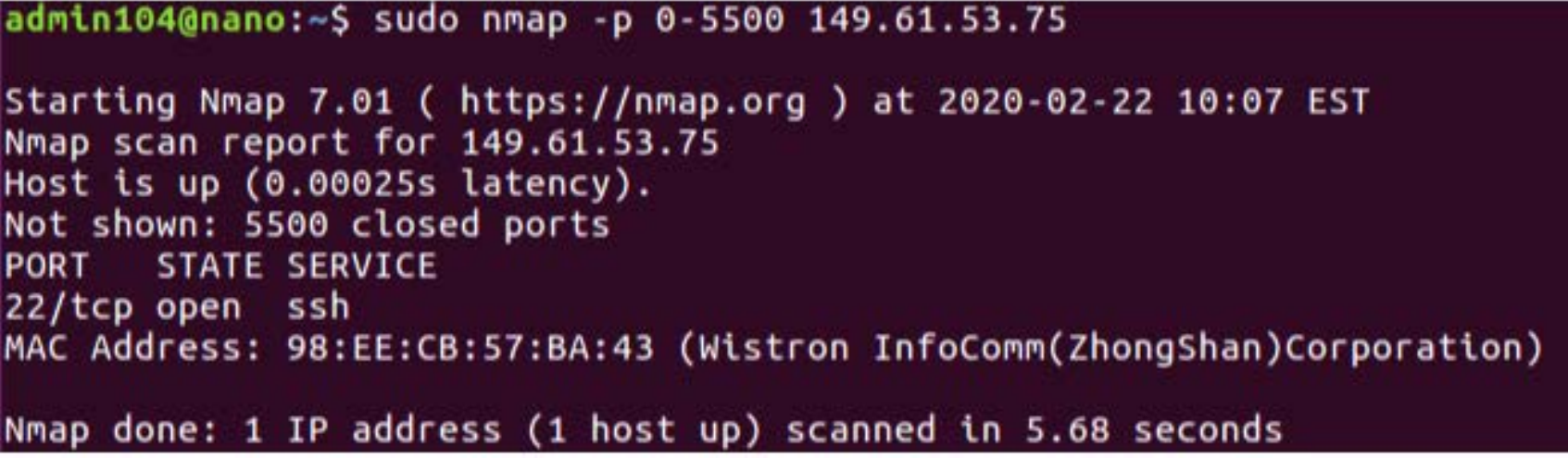}
    \caption{Port Scan Attack Without SDP}
    \label{fig:scan1}
\end{figure}

\begin{figure}[h]
        \centering
        \includegraphics[width=0.5\textwidth]{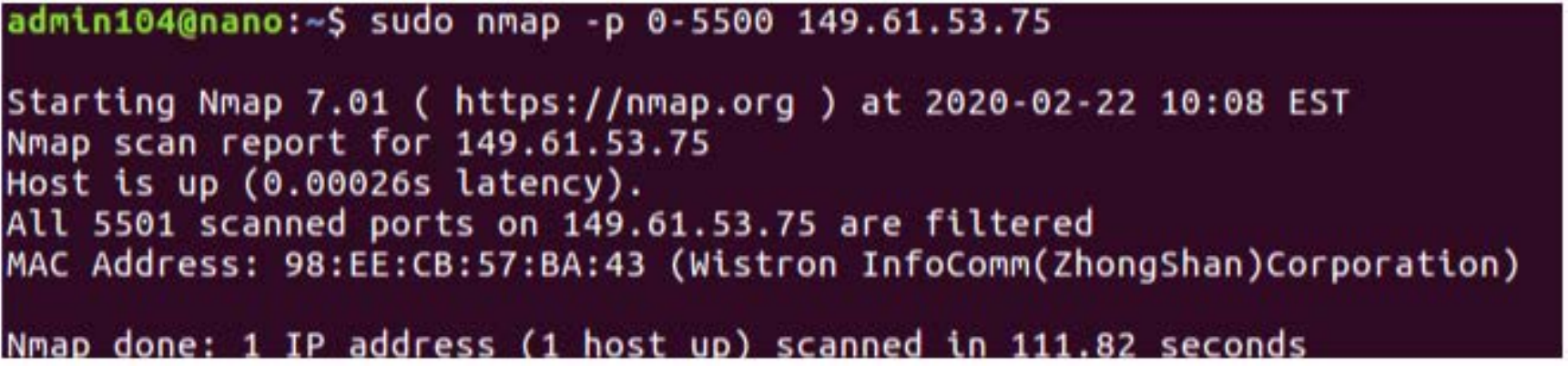}
        \caption{Port Scan Attack With SDP}
        \label{fig:scan2}
\end{figure}

A port scan attack was also performed to verify the gateway configuration, shown in Figure~\ref{fig:PS}. For this purpose, we utilized the free nmap utility tool to perform the attack with and without SDP protection as displayed in Figure~\ref{fig:scan1} and Figure~\ref{fig:scan2} respectively. The SDP client successfully uses port 5000 to authenticate to the controller and port 4444 to access the service after authentication, however, the port scan which was performed by the attack UE shows ports 0-5500 as being closed when SDP is up. This verifies the behavior of the gateway to block all traffic unless from authenticated sources. However, when the same attack was launched without SDP, the results show the TCP ssh connection as open and available. This further confirms the effectiveness of SDP framework in blackening the network from unauthorized users.


\begin{figure}[h]
    \centering
    \includegraphics[width=.5\textwidth]{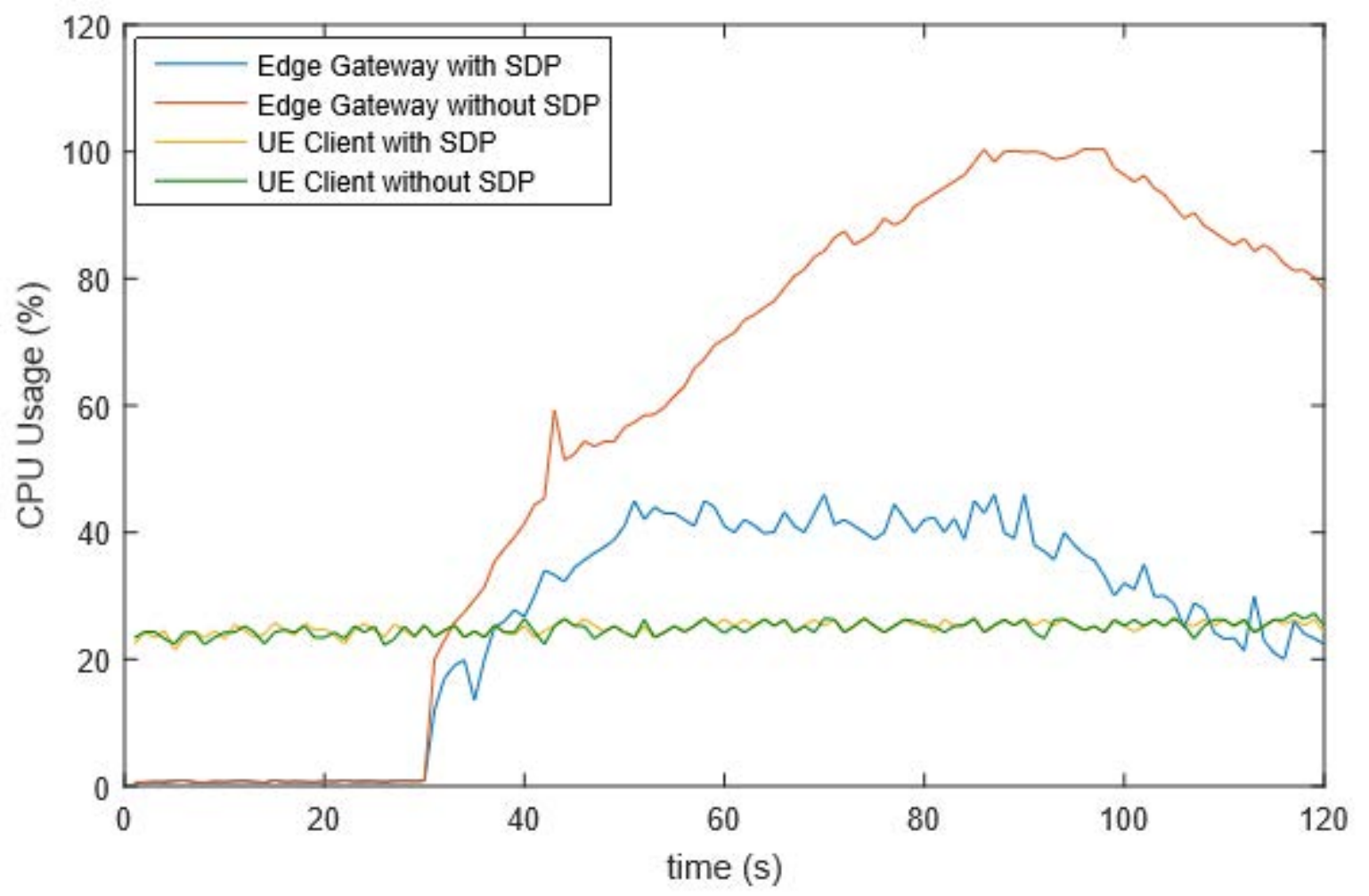}
    \caption{CPU Usage under DOS Attack}
    \label{fig:cpu}
\end{figure}

A CPU usage analysis is also performed and the results of which are shown in Figure~\ref{fig:cpu}. The figure shows the load variation under an attack with and without SDP at the gateway and the client. The attack is launched at 30s, shortly thereafter CPU usage increases. Without SDP, CPU usage at the gateway hits and even exceeds 100\%. On the contrary, with SDP enabled, the CPU usage increases slowly until it reaches 40\% and then reduces thereafter. For the UE client, the CPU usage only accounts for the processing power needed to run the UE and the SDP initiating host module on the physical machine emulating the UE. Thus, the Client's CPU usage is uncorrelated with the attack as displayed by the green and yellow curves in Figure~\ref{fig:cpu}.

\section{Conclusion}

This paper demonstrated the implementation of a combined MEC-SDP combined architecture for the LTE Core network. The security threats prevalent in MEC for LTE were discussed. The framework which was proposed to mitigate these threats was SDP. The architecture was described in detail followed by the implementation and testing. The results were then evaluated. The proposed SDP was capable of protecting cloud resources at the edge. SDP was able to drop the attack and showed a reduced CPU load compared to the unsecured case. A port scanning attack was also performed to verify the 'darkening' of services under the SDP framework. The attacker was not aware of ports and services, and should they be aware, will not have access to them without presenting a valid SPA message.

This study and its results open new challenges for future work. Firstly, the architecture explored only an edge gateway model, however, for additional security, a cloud gateway for each service can be introduced. In this way, services are only accessible by UEs apart of the LTE network that can interface with the edge controller. Secondly, edge services such as the aforementioned edge location services can be secured via SDP. This maintains a true zero trust architecture design as no location in the network is considered trusted.

\section*{Acknowledgements}

The authors would like to thank Juanita Koilpillai, Dan Bailey, and Todd McAnally from  Waverley Labs, for their valuable support. This work was made possible by grant \# IRCC-2020-003. The findings achieved herein are
solely the responsibility of the authors.

\bibliographystyle{IEEEtran}
\bibliography{ref}

\begin{IEEEbiography}[{\includegraphics[width=1in,height=5in,clip,keepaspectratio]{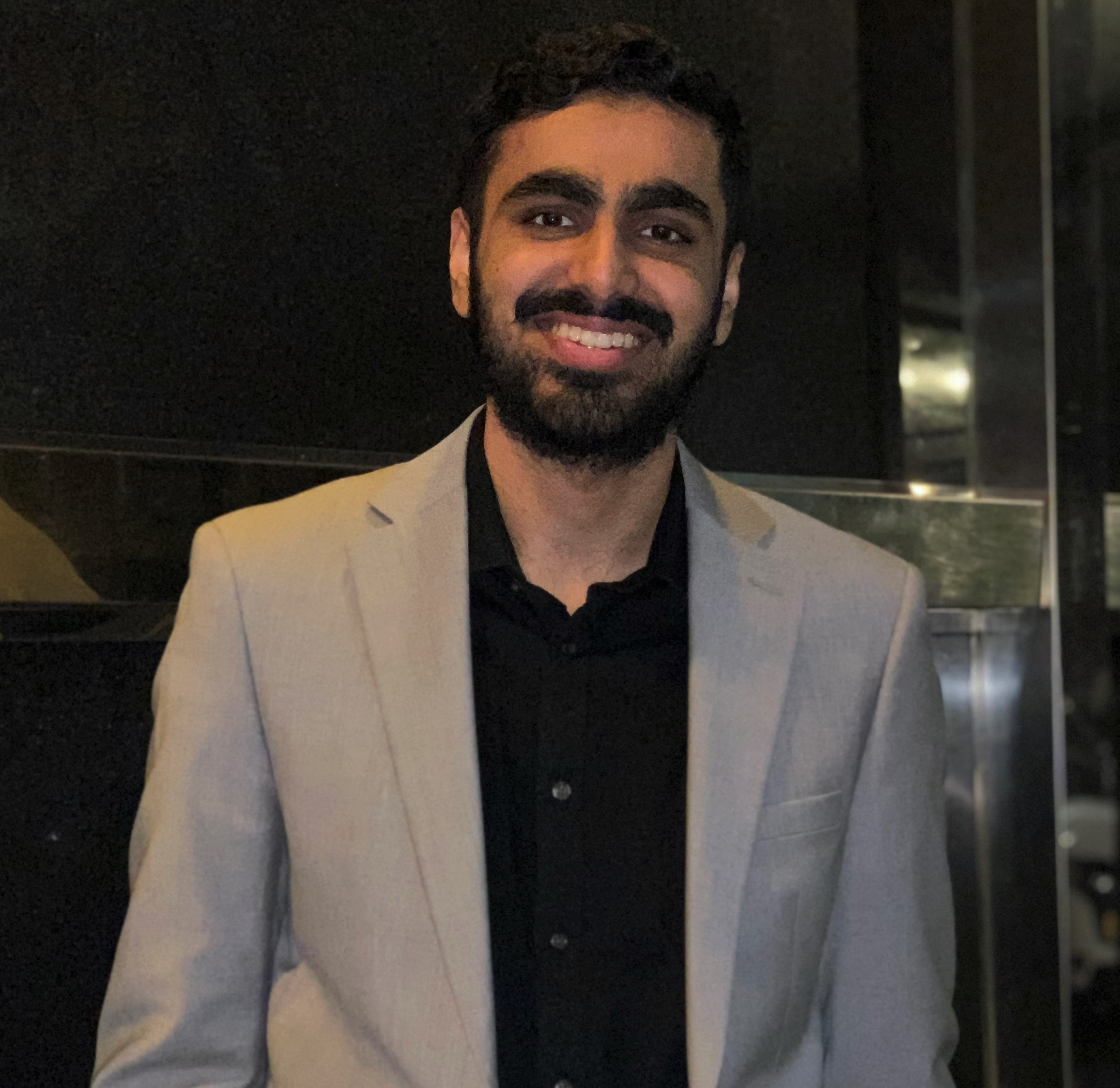}}]{Jaspreet Singh} is a M.Sc. Student at Manhattan College as well as a blockchain developer for Accenture LLP, certified in multiple blockchain platforms such as Corda and Hyperledger Fabric. Jaspreet focuses primarily on architecting and implementing of server-less cloud computer applications for AWS and GCP. He received his B.S with Applied Mathematics Concentration in computer engineering as well as M.S in computer engineering at Manhattan College. The focus of his master's research was on Cloud and Edge Computing security.
\end{IEEEbiography}

\begin{IEEEbiography}[{\includegraphics[width=1in,height=5in,clip,keepaspectratio]{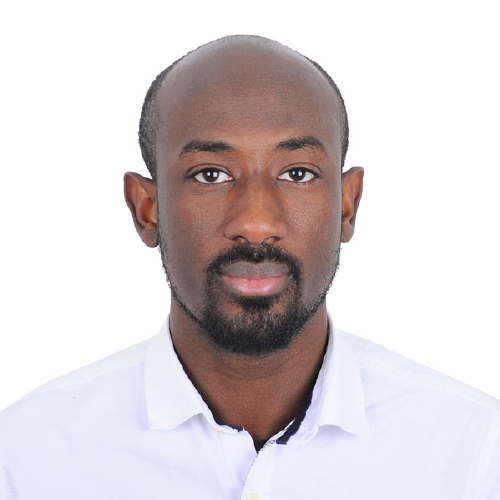}}]{Yahuza Bello} received the BSc degrees in electronics and communications engineering from the Arab Academy for Science, Technology and Maritime Transport (AASTMT), Egypt, in 2014. He is currently working toward the MSC degree in computer engineering at Manhattan College, Riverdale, New York, United State of America. His research interests include software defined networks, network function virtualization, resource allocation, cloud and edge computing security. 
\end{IEEEbiography}

\begin{IEEEbiography}[{\includegraphics[width=1in,height=5in,clip,keepaspectratio]{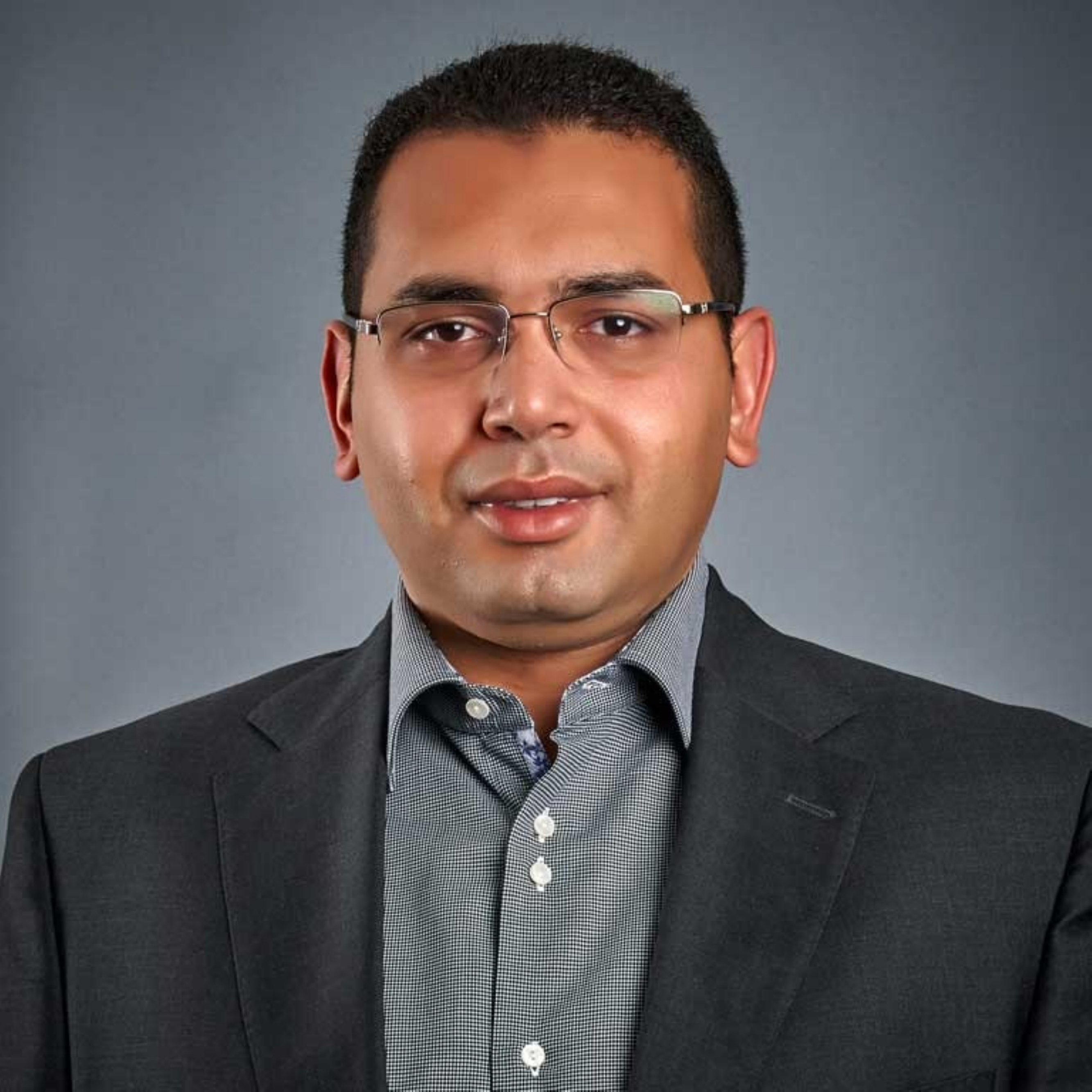}}]{Dr. Ahmed Refaey Hussein} (S’03-M’07-SM’15) is an Assistant Professor at Manhattan College as well as an adjunct research professor at Western University. Previously, Dr. Hussein’s positions included: a Sr. Embedded Systems Architect, R\& D group, Mircom Technologies Ltd from 2013-2016; and as a Postdoctoral Fellow at ECE department, Western University from 2012-2013; and Professional Researcher at the LRTS lab, Laval University in the field of wireless communications hardware implementations from 2007-2011. Prior to joining Laval University, Dr. Hussein was a System/ Core Network Engineer leading a team of junior engineers and technicians in the telecom field in the three prominent companies of Fujitsu, Vodafone, and Alcatel-Lucent. Dr. Hussein received his B.Sc. and M.Sc. degrees from Alexandria University, Egypt in 2003 and 2005, respectively; and Ph.D. degree from Laval University, Quebec, Canada in 2011. Dr. Hussein is the author and Co-author of more than 45 technical papers, 1 patent granted, and 3 patent applications addressing his research activities.
\end{IEEEbiography}

\begin{IEEEbiography}[{\includegraphics[width=1in,height=5in,keepaspectratio]{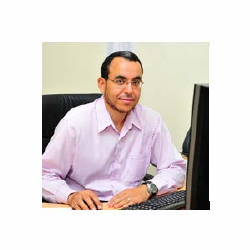}}]{Dr. Amr Mohamed} (S’00-M’06-SM’14) received his M.S. and Ph.D. degrees in electrical and computer engineering from the University of British Columbia, Vancouver, Canada, in 2001 and 2006, respectively. He worked as an advisory IT specialist at the IBM Innovation Centre in Vancouver from 1998 to 2007, taking a leadership role in systems development for vertical industries. He is currently a professor in the College of Engineering at Qatar University and the Director of the Cisco Regional Academy. He has over 25 years of experience in wireless networking research and industrial systems development. He holds three awards from IBM Canada for his achievements and leadership, and four best paper awards from IEEE conferences. His research interests include wireless networking and edge computing for IoT applications. He has authored or co-authored over 200 refereed journal and conference papers, textbooks, and book chapters in reputable international journals, and conferences. He is serving as a technical editor for two international journals and has served as a technical program committee (TPC) co-chair for many IEEE conferences and workshops.
\end{IEEEbiography}
\end{document}